\documentclass[aps,prx,superscriptaddress,twocolumn,longbibliography]{revtex4-1}
\usepackage{units}
\usepackage{amsmath}
\usepackage{amsthm}
\usepackage{amssymb}
\usepackage{graphicx}
\usepackage{color}
\usepackage{bbold}
\usepackage[colorlinks=true, urlcolor=blue, citecolor=blue,linkcolor=blue,citebordercolor={1 0 0},linkbordercolor={0 0 1}]{hyperref}

\graphicspath{{./images/},{./imagesAppendix/}}

\theoremstyle{plain}
\newtheorem{thm}{\protect\theoremname}
\theoremstyle{plain}
\newtheorem{fact}[thm]{\protect\factname}
\ifx\proof\undefined
\newenvironment{proof}[1][\protect\proofname]{\par
	\normalfont\topsep6\p@\@plus6\p@\relax
	\trivlist
	\itemindent\parindent
	\item[\hskip\labelsep\scshape #1]\ignorespaces
}{%
	\endtrivlist\@endpefalse
}
\providecommand{\proofname}{Proof}
\fi
\theoremstyle{plain}
\newtheorem{lem}[thm]{\protect\lemmaname}
\theoremstyle{remark}

\newcommand{\bra}[1]{\langle #1|}
\newcommand{\ket}[1]{|#1 \rangle}
\makeatother
\newcommand{\braket}[2]{\langle #1 \vert #2 \rangle}
\newcommand{\abs}[1]{\left|#1\right|}
\newcommand{\idg}[1]{{\bfseries #1)}}

\newcommand\numberthis{\addtocounter{equation}{1}\tag{\theequation}}
\providecommand{\factname}{Fact}
\providecommand{\theoremname}{Theorem}
\providecommand{\claimname}{Claim}
\providecommand{\lemmaname}{Lemma}
\providecommand{\definitionname}{Definition}

\theoremstyle{definition}
\newtheorem{defn}[thm]{\protect\definitionname}

\newcommand{\revA}[1]{{#1}}
\newcommand{\revB}[1]{{#1}}

\newcommand{\subfigimg}[3][,]{%
	\setbox1=\hbox{\includegraphics[#1]{#3}}% Store image in box
	\leavevmode\rlap{\usebox1}% Print image
	\rlap{\hspace*{2pt}\raisebox{\dimexpr\ht1-0.5\baselineskip}{{\bfseries \large\textsf{#2}}}}% Print label
	\phantom{\usebox1}% Insert appropriate spcing
}

\begin{document}
\title{Quantum Assisted Simulator}
\author{Kishor Bharti}
\email{kishor.bharti1@gmail.com}
\affiliation{Centre for Quantum Technologies, National University of Singapore 117543, Singapore}
\author{Tobias Haug}
\affiliation{Centre for Quantum Technologies, National University of Singapore 117543, Singapore}
\begin{abstract}
Quantum simulation can help us study poorly understood topics such as high-temperature superconductivity and drug design. However, existing quantum simulation algorithms for current quantum computers often have drawbacks that impede their application.
Here, we provide a novel hybrid quantum-classical algorithm for simulating the dynamics of quantum systems. Our approach takes the Ansatz wavefunction as a linear combination of quantum states. The quantum states are fixed, and the combination parameters are variationally adjusted. Unlike existing variational quantum simulation algorithms, our algorithm does not require any classical-quantum feedback loop and by construction bypasses the barren plateau problem. Moreover, our algorithm does not require any complicated measurements such as the Hadamard test. The entire framework is compatible with existing experimental capabilities and thus can be implemented immediately.
\end{abstract}

\maketitle
\section{Introduction}
%\noindent {\em Introduction.---} 
The near-term success of the second quantum revolution critically depends
on the practical applications of noisy intermediate-scale quantum (NISQ)~\cite{preskill2018quantum,deutsch2020harnessing,bharti2021noisy}
devices. Though experimental demonstration of ``quantum supremacy''
has induced widespread hope~\cite{arute2019quantum}, the essential question regarding how to
translate such breakthroughs into quantum advantages for practical
use-cases remains unsolved. The search for the ``killer app'' for NISQ devices continues, with potential areas of application being solid-state physics, quantum
chemistry and combinatorial optimization. Most of the problems from the aforementioned areas can be
mapped are Hamiltonian ground state problem and the simulation of
quantum dynamics. Variational classical simulation (VCS) techniques
have been suggested to handle static problems in estimating the
ground state and ground state energy of a Hamiltonian, as well as
dynamical problems in simulating the time evolution (real as well
as imaginary) of quantum systems. However, for problems involving
exponentially large Hilbert spaces, VCS in general fails to provide the desired solution. 

The canonical NISQ era algorithm for approximating the ground state of a Hamiltonian is the variational quantum eigensolver (VQE) \cite{peruzzo2014variational,mcclean2016theory,kandala2017hardware,farhi2014quantum,farhi2016quantum,harrow2017quantum, farhi2016quantum,mcardle2020quantum,endo2020hybrid}. 
The VQE is a hybrid quantum-classical algorithm that employs a classical optimizer to tune the parameters of a parameterized quantum circuit using measurements performed on the quantum device.
The classical optimization landscape corresponding to VQE is highly non-convex and in general uncharacterized, rendering any proper theoretical study difficult~\cite{bittel2021training}. Moreover, the recent results on the appearance of the barren
plateau as the hardware noise, amount of entanglement or number of qubits increase, has led to genuine concerns regarding the fate of VQE \cite{mcclean2018barren,huang2019near,sharma2020trainability,cerezo2020cost,wang2020noise,marrero2020entanglement}.
To tackle the existing challenges in VQE, the quantum assisted eigensolver (QAE) \cite{bharti2020quantum} and iterative quantum assisted eigensolver (IQAE) \cite{bharti2020iterative} have been recently proposed in the literature. The classical optimization program of algorithms is a quadratically constrained quadratic program with single equality constraint, which is a well characterized optimization program. In particular, the IQAE algorithm provides a systematic path to build Ansatz, bypasses the barren plateau problem
and can be efficiently implemented on existing hardware.

The broader task of simulating quantum dynamics is challenging as the Hilbert space dimension increases exponentially, which poses a considerable bottleneck to the study and design of new drugs, catalysts, and materials. A universal quantum computer, with millions of qubits with noise levels beneath a critical threshold offers a possibility to simulate the dynamics.
Simulation algorithms such as Trotterization usually require many quantum gates, which most likely would require the use of fault tolerant quantum computers to implement~\cite{poulin2014trotter}.
However, most likely fault-tolerant quantum computers will not be available in the near feature.
Thus, to harness the potential of NISQ hardware, variational quantum simulation (VQS) algorithms have been suggested in literature~\cite{li2017efficient,mcardle2019variational,yuan2019theory}. The algorithm is hybrid quantum-classical in nature
and utilizes dynamical variational principles to update the parameters
of a parametric quantum circuit, such that the Ansatz evolution emulates quantum evolution. \revA{However, the VQS algorithm shares the issues faced by VQE as it may suffer from the barren plateau problem as well.} Moreover, it requires complicated
measurements involving controlled unitaries, which may not be viable
in the NISQ era. The classic-quantum feedback loop requires extensive waiting times in the queues of cloud based quantum computers which slows down the algorithm on existing quantum hardware. Furthermore, the Ansatz if VQS is often not chosen in a systematic way and typically requires adjustable parameters to be real-valued~\cite{yuan2019theory}.

In this work, we provide a novel hybrid quantum-classical algorithm for simulating the dynamics of
quantum systems. We refer to our algorithm as quantum assisted simulator (QAS)
algorithm. 
The Hamiltonian is assumed to be a linear combination
of unitaries, and the Ansatz is a linear combination of quantum
states. The combination coefficients are complex-valued, in general.
Our algorithm can perform both real and imaginary time evolution of Hamiltonians.
Unlike existing variational quantum simulation algorithms,
our algorithm does not mandate any classical-quantum feedback loop, which speeds up computations on current cloud-based quantum computers.
By construction, the algorithm circumvents the barren plateau problem.
Our algorithm does not demand any complicated measurement involving controlled unitaries and can be run by simple measurements of Pauli strings. The entire framework is compatible with existing
experimental capabilities and thus can be implemented immediately.

\begin{figure}[htbp]
	\centering
	\subfigimg[width=0.49\textwidth]{}{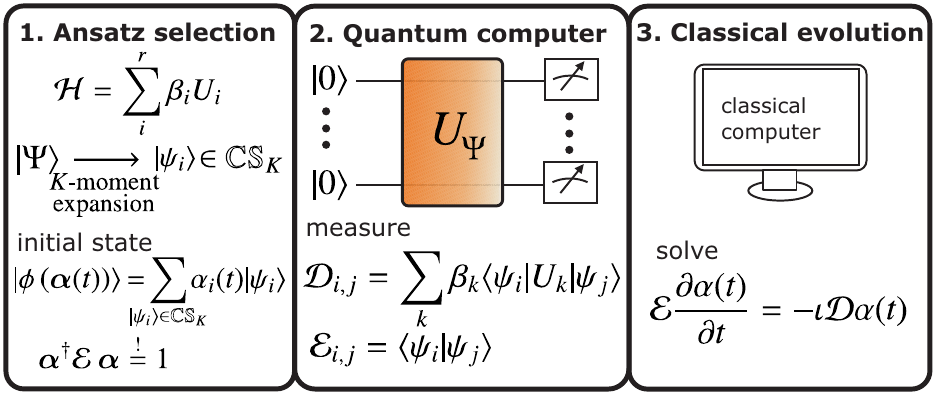}
	\caption{The Quantum Assisted Simulator (QAS) algorithm consists of three steps. The first step involves the selection of the Ansatz. The Ansatz is represented as a linear combination of cumulative $K$-moment states (see Definition \ref{def:cumulant_states}), for some choice of non-negative integer $K$. The second step computes overlap matrices on a quantum computer, which can be performed efficiently without using complicated measurement techniques (see Appendix~\ref{sec: Measurements}).  After the overlap matrices have been computed, the differential equation corresponding to Eq.\eqref{eq:Evolution_Frenkel_Complex} is solved on a classical computer. Unlike existing variational quantum simulation algorithms, QAS does not mandate any classical-quantum feedback loop.}
	\label{QASComic}
\end{figure}

%\medskip
%{\noindent {\em QAS Algorithm---}}
\section{QAS Algorithm}
The time evolution of a closed system, represented by the time-dependent quantum state $\vert\psi(t)\rangle$ for Hamiltonian $H$ is given by
\begin{equation}
\frac{d\vert\psi(t)\rangle}{dt}=-\iota H\vert\psi(t)\rangle.\label{eq:Schrodinger}
\end{equation}
Let us consider the Hamiltonian $H$ as a linear combination of $r$
unitaries
\begin{equation}
H=\sum_{i=1}^{r}\beta_{i}U_{i},\label{eq:Ham_LCU_Frenkel}
\end{equation}
where the combination coefficients $\beta_{i}\in\mathbb{C}$ and the
$N$-qubit unitaries $U_{i}\in \text{SU}\left(2^{N}\equiv \mathcal{N}\right)$, for $i\in\left\{ 1,2,\cdots,r\right\} $.  Moreover, each unitary acts non-trivially on at most $\mathcal{O}\left(poly\left(logN\right)\right)$ qubits. If the unitaries in Eq.\eqref{eq:Ham_LCU_Frenkel} are tensored Pauli matrices, we do not need the aforementioned  $\mathcal{O}\left(poly\left(logN\right)\right)$ constraint.
Let us consider the Ansatz state as time dependent linear combination
of $m$ quantum states $\left\{ \vert\psi_{i}\rangle\right\} _{i}$
\begin{equation}
\vert\phi\left(\boldsymbol{\alpha}(t)\right)\rangle=\sum_{i=0}^{m-1}\alpha_{i}(t)\vert\psi_{i}\rangle,\label{eq:Ansatz}
\end{equation}
for $\alpha_{i}(t)\in\mathbb{C}$. Normalization of the Ansatz wavefunction $\braket{\phi(\alpha)}{\phi(\alpha)}=1$ is achieved by demanding
\begin{equation}\label{Enorm}
\boldsymbol{\alpha}^\dagger \mathcal{E}\boldsymbol{\alpha}=1,
\end{equation}
where
\begin{equation}
\mathcal{E}_{i,j}=\langle\psi_{i}\vert\psi_{j}\rangle. \label{eq:Overlap_E_2}
\end{equation}
Using Dirac and Frenkel variational
principle~\cite{dirac1930note,frenkel1934wave}, we get
\begin{equation}
\langle\delta\phi\left(\boldsymbol{\alpha}(t)\right)\vert\left(\frac{d}{dt}+\iota H\right)\vert\phi\left(\boldsymbol{\alpha}\right)\rangle=0,\label{eq:Frenkel}
\end{equation}
where 
\begin{equation}
\langle\delta\phi\left(\boldsymbol{\alpha}(t)\right)\vert=\sum_{i}\frac{\partial\langle\phi\left(\boldsymbol{\alpha}(t)\right)\vert}{\partial\alpha_{i}^*}\delta\alpha_{i}^*(t)\label{eq:Frenkel_variation}
\end{equation}
The evolution of $\boldsymbol{\alpha}(t)$ can be solved to be
\begin{equation}
\sum_{j}\mathcal{E}_{i,j}\dot{\alpha_{j}}=-\iota\mathcal{C}_{i},\label{eq:Evolution_Frenkel_01}
\end{equation}
where $\mathcal{E}$ and $\mathcal{C}$ are 
\begin{equation}
\mathcal{E}_{i,j}=\frac{\partial\langle\phi\left(\boldsymbol{\alpha}(t)\right)\vert}{\partial\alpha_{i}^*}\frac{\partial\vert\phi\left(\boldsymbol{\alpha}(t)\right)\rangle}{\partial\alpha_{j}},\label{eq:Overlap_E}
\end{equation}
\begin{equation}
\mathcal{C}_{i}=\sum_{j}\alpha_{j}(t)\frac{\partial\langle\phi\left(\boldsymbol{\alpha}(t)\right)\vert}{\partial\alpha_{i}^*}H\vert\psi_{j}\rangle.\label{eq:Overlap_C}
\end{equation}
Using Eq.\ref{eq:Ham_LCU_Frenkel}, we define
\begin{equation}
\mathcal{D}_{i,j}=\sum_{k}\beta_{k}\langle\psi_{i}\vert U_{k}\vert\psi_{j}\rangle.\label{eq:Overlap_D}
\end{equation}
Notice that 
\begin{equation}
\frac{\partial\vert\phi\left(\boldsymbol{\alpha}\right)\rangle}{\partial\alpha_{j}}=\frac{\partial\sum_{i=0}^{m-1}\alpha_{i}(t)\vert\psi_{i}\rangle}{\partial\alpha_{j}}\label{eq:derivative_01-1}
\end{equation}
\begin{equation}
\implies\frac{\partial\vert\phi\left(\boldsymbol{\alpha}\right)\rangle}{\partial\alpha_{j}}=\vert\psi_{j}\rangle.\label{eq:derivative_02-1}
\end{equation}
Thus, we identify $\mathcal{E}$ of Eq.\eqref{eq:Overlap_E} and Eq.\eqref{eq:Overlap_E_2} as being identical and
\begin{equation}
\mathcal{C}_{i}=\sum_{j}\mathcal{D}_{i,j}\alpha_{j}(t).\label{eq:Overlap_C_2}
\end{equation}
Finally, we have
\begin{equation}
\mathcal{E}\frac{\partial\boldsymbol{\alpha}(t)}{\partial t}=-\iota\mathcal{D}\boldsymbol{\alpha}(t).\label{eq:Evolution_Frenkel_Complex}
\end{equation}
Thus, one can solve for $\dot{\alpha_{j}}(t)$ and hence update the
parameters as 
\begin{equation}
\alpha_{j}(t+\delta t)=\alpha_{j}(t)+\dot{\alpha_{j}}\delta t,\label{eq:Time_evolution_alpha_1-1}
\end{equation}
for an evolution corresponding to time $\delta t$. It is easy to show that the evolution keeps $\boldsymbol{\alpha}(t)$ normalized according to Eq.\eqref{Enorm}. Similarly, we can define the QAS algorithm for imaginary time evolution by substituting $t$ with $-\iota\tau$ in Eq.\eqref{eq:Schrodinger} (see Appendix~\ref{sec: ITE}).

For pedagogical reasons, we use the
notion of $K$-moment states and cumulative $K$-moment states \cite{bharti2020iterative}.
\begin{defn}  \label{def:cumulant_states}
Given a set of unitaries $\mathbb{U}\equiv\left\{ U_{i}\right\} _{i=1}^{r}$,
a positive integer $K $ and some quantum state $\vert\psi\rangle,$
$K$-moment states is the set of quantum states of the form $\left\{ U_{i_{K}}\cdots U_{i_{2}}U_{i_{1}}\vert\psi\rangle\right\} _{i}$
for  $U_{i_{l}}\in\mathbb{U}.$ We denote the aforementioned set by $\mathbb{S}_{K}$. The singleton set $\left\{ \vert\psi\rangle\right\} $
will be referred to as the $0$-moment state (denoted by $\mathbb{S}_{0}$.) The cumulative
$K$-moment states $\mathbb{CS}_{K}$ is defined to be $\mathbb{CS}_{K}\equiv\cup_{j=0}^{K}\mathbb{S}_{j}$.
\end{defn}
For example, the set of $1$-moment states is given by $\left\{ U_{i}\vert\psi\rangle\right\} _{i=1}^{r}$, 
for a given initial state $\vert\psi\rangle$ and the unitaries $\left\{ U_{i}\right\} _{i=1}^{r}$
which define the Hamiltonian $H$. The set of cumulative  $1$-moment states is given by $\mathbb{CS}_{1}=\{\vert \psi \rangle\} \cup \left\{ U_{i}\vert\psi\rangle\right\} _{i=1}^{r}$, and the set describing cumulative $K$-moment states is $\mathbb{CS}_{K}=\{\vert \psi \rangle\} \cup \left\{ U_{i_1}\vert\psi\rangle\right\} _{i_1=1}^{r} \cup \dots \cup \left\{ U_{i_K}\dots U_{i_1}\vert\psi\rangle\right\} _{i_1=1,\dots,i_K=1}^{r}$.

\revA{The accuracy of QAS improves with increasing moment $K$ and number of measured overlaps. The $K$-moment states are generated 
by the action of the product of unitaries on a fixed quantum state. Thus, when the set of unitaries used to generate the $K$-moment states attains closure under multiplication, the set of cumulative $K$-moment states also closes. 
In such cases, the cumulative $K$-moment states do not increase any further as we increase $K$. 
Here, the Lie group structure of the unitaries defining the Hamiltonian plays a crucial role for what $K$ we see closure.}

The QAS algorithm involves three steps (see Fig.\ref{QASComic} for pictorial synopsis).
\begin{enumerate}
\item Ansatz selection
\item Calculation of the overlap matrices $\left(\mathcal{D}\text{ and }\mathcal{E}\right)$
on a quantum computer
\item Solving the differential equation using equations
\ref{eq:Evolution_Frenkel_Complex} on a classical computer
\end{enumerate}
The first step of the QAS algorithm is crucial and heavily determines
the accuracy of the algorithm. We consider our Ansatz to be a linear
combination of cumulative $K$-moment states, i.e.,
\begin{equation}
\vert\phi\left(\boldsymbol{\alpha}(t)\right)\rangle=\sum_{\vert\psi_{i}\rangle\in\mathbb{CS}_{K}}\alpha_{i}(t)\vert\psi_{i}\rangle,\label{eq:K-moment Ansatz}
\end{equation}
where $\alpha_{i}(t)\in\mathbb{C}.$ 
For the given selection of the Ansatz
in Eq.\ref{eq:K-moment Ansatz}, the second step of the QAS algorithm
involves computing the overlap matrices. The matrix elements of $\mathcal{D}\text{ and }\mathcal{E}$
can be estimated efficiently on a quantum computer, using the techniques
from Mitarai et al., without any complicated measurement such as the
Hadamard test \cite{mitarai2019methodology}. If the Hamiltonian is a linear combination of tensored
Pauli operators, the elements can be directly inferred from measurements
in the corresponding computational basis as we further discuss later on as well as in Appendix~\ref{sec: Measurements}.
Once the overlap matrices have been computed, the job of the quantum
computer is over. For the third and final step of the QAS algorithm we
solve Eq.~\ref{eq:Evolution_Frenkel_Complex} on a classical computer. If the desired accuracy has not been attained,
one can re-run the whole algorithm for an increased choice of $K$.

%\medskip
%{\noindent {\em Justification for the Ansatz---}}
\section{Justification for the Ansatz}
We now proceed to provide a small justification for the choice of the
Ansatz in the QAS algorithm. Suppose the initial state ($0$-moment
state) is $\vert\psi\rangle$. Starting with $\vert\psi\rangle$,
if one applies $\exp\left(-iHt\right)$ for some $t\geq0$, the
evolved state is given by 
\begin{equation}
\vert\gamma\rangle=e^{-iHt}\vert\psi\rangle.\label{eq:ITE_state_1}
\end{equation}
Using $e^{-iHt}=\sum_{p=0}^{\infty}\frac{\left(-iHt\right)^{p}}{p!}$,
we get
\begin{equation}
\vert\gamma\rangle=\sum_{p=0}^{\infty}\frac{\left(-iHt\right)^{p}}{p!}\vert\psi\rangle.\label{eq:ITE_state_expanded}
\end{equation}
 Let us define the operator 
\begin{equation}
\mathcal{O}^{K}\equiv\sum_{p=0}^{K}\frac{\left(-iHt\right)^{p}}{p!},\label{eq:K-Approx_Operator-1}
\end{equation}
for $K\geq0.$ Notice that $\mathcal{O}^{K}$ corresponds to the sum
of first $K$ terms of $e^{-iHt}.$ Using $\mathcal{O}^{K},$ we proceed
to define
\begin{equation}
\vert\gamma_{K}\rangle\equiv\frac{\sum_{p=0}^{K}\frac{\left(-iHt\right)^{p}}{p!}\vert\psi\rangle}{\sqrt{\langle\psi\vert\left(\sum_{p=0}^{K}\frac{\left(-iHt\right)^{p}}{p!}\right)^{2}\vert\psi\rangle}}.\label{eq:Evolved_state_ITE_K-Approx-1}
\end{equation}
For $K\rightarrow\infty$, $\vert\gamma_{K}\rangle\rightarrow\vert\gamma\rangle.$
Using the expression for Hamiltonian as linear combination of unitary,
it is easy to see that $\vert\gamma_{K}\rangle$ can be written as
linear combination of cumulative $K$-moment states, i.e,
\[
\vert\gamma_{K}\rangle=\sum_{\vert u_{i}\rangle\in\mathbb{CS}_{K}}\alpha_{i}\vert u_{i}\rangle
\]
where the combination coefficients $\alpha_{i}\in\mathbb{C}.$ The
aforementioned arguments justify the choice of Ansatz as linear combination
of cumulative $K$-moment states. 
\revA{ Our Ansatz is based on the Krylov subspace idea. For a scalar $\tau$, a $\mathcal{N}\times \mathcal{N}$ matrix $A$ and
a $\mathcal{N}$-dimensional vector $v$, the action of the matrix exponential operator
$\exp\left(\tau A\right)$ on $v$ can be approximated as~\cite{lanczos1950iteration,saad1992analysis,seki2020quantum,motta2020determining}
\begin{equation}
\exp\left(\tau A\right)v\approx p_{K-1}\left(\tau A\right)v,\label{eq:approx_Krylov}
\end{equation}
where $p_{K-1}$ is a $K-1$ degree polynomial. The approximation
in Eq.\eqref{eq:approx_Krylov} is an element of the Krylov subspace,
\begin{equation}
Kr_{K-1}\equiv span\left\{ v,Av,\cdots,A^{K-1}v\right\} .\label{eq:Krylov_1}
\end{equation}
Thus, the problem of approximating $\exp\left(\tau A\right)v$ can
be recast as finding an element from $Kr_{K}.$ Note that the approximation
in Eq.\eqref{eq:approx_Krylov} becomes exact when $K-1=rank(A).$
In our case, we can identify $v$ with the initial state $\vert\psi\rangle$, $\tau$ with $-\iota t$
and $A$ with the Hamiltonian $H$. 
\revB{Thus, one could implement the Krylov subspace ansatz directly with QAS by using the ansatz space $\mathbb{CS}_{K}=\{\ket{\psi},H\ket{\psi},\cdots,H^{K-1}\ket{\psi}\}$. However, here we would have to estimate overlaps involving $H^k$, which is a challenge for NISQ computers. Importance sampling has been proposed to estimate $H^k$~\cite{mcclean2020decoding}, however this may require more measurements than current NISQ computers can handle. 
Our cumulative $K$-moment states as shown in Definition \ref{def:cumulant_states} subsumes the Krylov subspace, however in contrast can be measured in a NISQ friendly way as we show below.}
}

\revB{QAS requires the measurement of overlaps. For commonly used Hamiltonians, the unitaries $U_i$ in Eq.\eqref{eq:Ham_LCU_Frenkel} are Pauli strings $P_i=\bigotimes_{j=1}^N\boldsymbol{\sigma}_j$, where $\boldsymbol{\sigma}_j\in \{\mathbb{1},\sigma^x,\sigma^y,\sigma^z\}$. In this case, the overlaps can be easily measured on NISQ devices.
The matrix elements of $\mathcal{D}$ and $\mathcal{E}$ (Eqs.\ref{eq:Overlap_E},\ref{eq:Overlap_D}) are written as $\bra{\psi}\prod_{q} P_{q}\ket{\psi}=a\bra{\psi}P'\ket{\psi}$, which can be rewritten as a single Pauli string $P'$ with a prefactor $a\in\{1,-1,i,-i\}$. Thus, the measurement of overlaps simply becomes the measurement of Pauli strings, which is efficient on current NISQ devices via a simple sampling task. We discuss further details on measurements of overlaps in the Appendix~\ref{sec: Measurements}.
}

%\medskip
%{\noindent {\em Examples---}}
\section{Examples}
\begin{figure}[htbp]
	\centering
	\subfigimg[width=0.24\textwidth]{a}{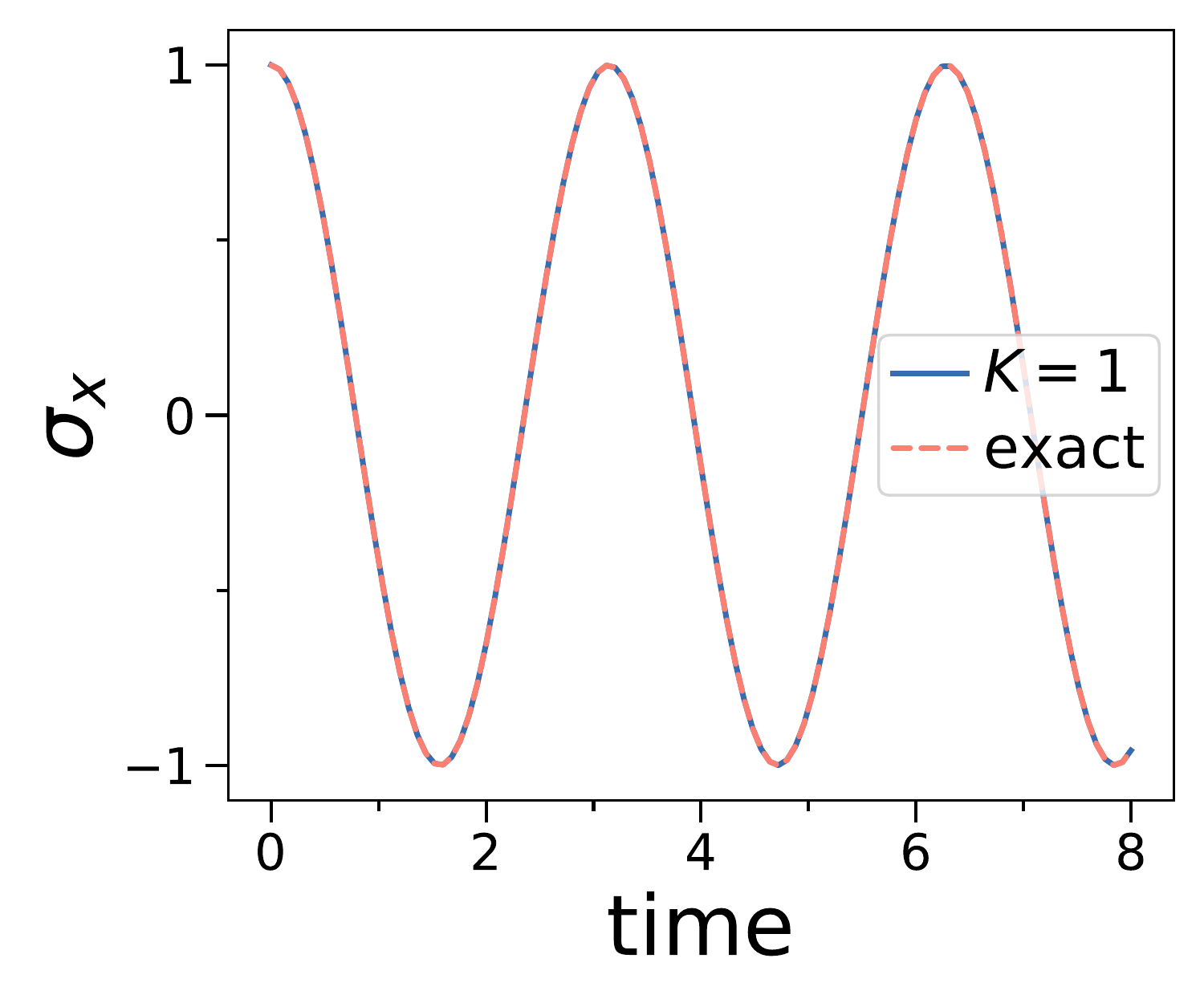}\hfill
	\subfigimg[width=0.24\textwidth]{b}{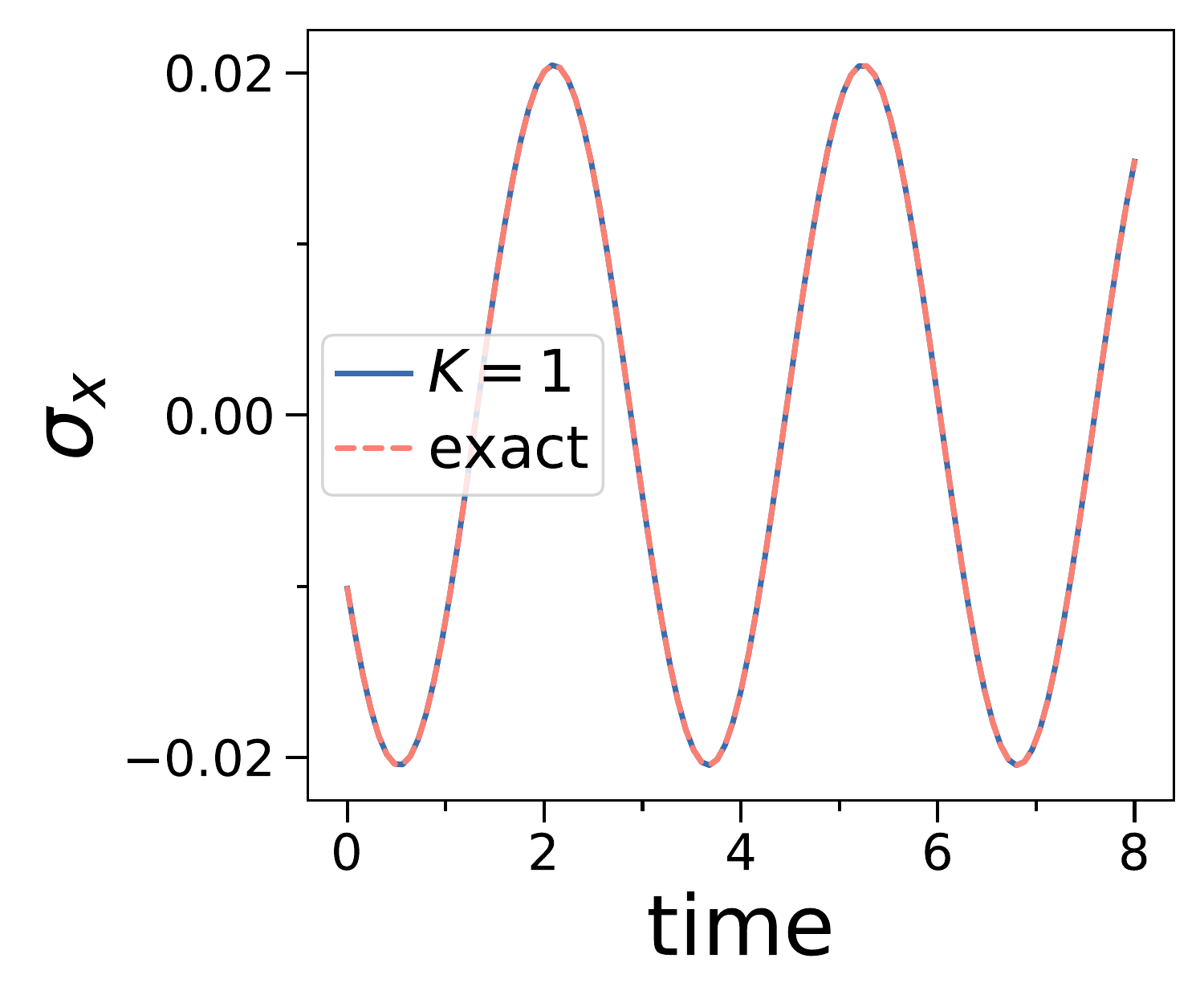}
	\caption{Time evolution with QAS. \idg{a} Evolution of expectation value of $\langle\sigma^x(t)\rangle$ for single qubit with Hamiltonian $H_\text{B}=\sigma^z$. We show the simulation using first moment expansion and the exact result. \idg{b} 
	Initial state is a deep circuit of $N=12$ qubits which for VQE is known to exhibit barren plateaus~\cite{mcclean2018barren}. It consists of $d=200$ layers composed of alternating single-qubit rotations around randomly chosen $x$-, $y$- or $z$-axis and C-phase gates arranged in a hardware efficient manner~\cite{mcclean2018barren}. We simulate the Hamiltonian $H_\text{bp}=\sigma^z_1\sigma^z_2$~\cite{mcclean2018barren}. Using QAS, we can simulate the exact evolution of the expectation value $\langle\sigma^x_1(t)\rangle$ already for the $K=1$ moment expansion.  }
	\label{Examples}
\end{figure}

We now show examples of the QAS algorithm applied to various Hamiltonians and Ansatz states.
First, in Fig.\ref{Examples}, we investigate two elementary examples. 
In Fig.\ref{Examples}a, we show the evolution with QAS of a single qubit for the Hamiltonian $H_\text{B}=\sigma^z$.
In step $1$ of the QAS algorithm, we choose the initial Ansatz state $\ket{\psi}=\ket{\psi_0}=\ket{+}$ (${0}$-moment state) $\mathbb{S}_0=\{\ket{\psi_0}\}$. Then, following Definition \ref{def:cumulant_states}, we use the set of unitaries that make up the Hamiltonian $\{\sigma^z\}$ and generate the $1$-moment states  $\mathbb{S}_1=\{\ket{\psi_1}\}$ with $\ket{\psi_1}=\sigma^z\ket{\psi_0}$. The union of $\mathbb{S}_1$ and $\mathbb{S}_0$ gives us the cumulative $1$-moment states $\mathbb{CS}_1=\{\ket{\psi_0},\ket{\psi_1}\}$. Higher orders $K$ of the moment expansion can be prepared by repeating this approach.
In step 2, we calculate the overlap matrices $\mathcal{D}_{n,m}=\bra{\psi_n}H_\text{B}\ket{\psi_m}$, $\mathcal{E}_{n,m}=\braket{\psi_n}{\psi_m}$ (Eq.\ref{eq:Overlap_E}, \ref{eq:Overlap_D}). 
Then, in step $3$, we choose initial state of evolution $\ket{\phi(t=0)}=\ket{\psi}$ with $\alpha_0(0)=1$ and $\alpha_1(0)=0$, and then evolve $\boldsymbol{\alpha}(t)$ to get the superposition state $\ket{\phi(t)}=\alpha_0(t)\ket{\psi_0}+\alpha_1(t)\ket{\psi_1}$. 
We show the evolution of the expectation value of the Pauli operator $\langle\sigma^x(t)\rangle=\bra{\phi(t)}\sigma^x\ket{\phi(t)}$. We find that the simulation with QAS and the exact evolution matches for the first moment expansion ($K=1$) \revA{as the set of $K$-moment states closes here already.}

\revB{Next, in Fig.\ref{Examples}b, our initial state for evolution is a deep quantum circuit $\ket{\Psi(\boldsymbol{\theta})}$ as studied in ~\cite{mcclean2018barren} to demonstrate the so-called barren plateau problem of variational quantum algorithms.
This Ansatz consists of $N$ qubits with $d\gg 1$ layers of unitaries, composed of alternating single-qubit rotations around randomly chosen $x$-, $y$- or $z$-axis parameterized with parameter $\boldsymbol{\theta}$ and control phase gates arranged in a hardware efficient manner.
This circuit suffers from the barren plateau problem where the variance of the gradients of the circuit in respect to an underlying cost function $H$ vanishes exponentially with the number of qubits $N$, i.e.  $\text{var}(\partial_{\boldsymbol{\theta}}\bra{\Psi(\boldsymbol{\theta})}H\ket{\Psi(\boldsymbol{\theta})})~\propto\exp(-N)$. 
For variational quantum algorithms, evolving these states is difficult as the gradients become exponentially small with increasing $N$. 
While optimizing the circuit parameters $\boldsymbol{\theta}$ is difficult, we show that we can simulate quantum dynamics with the very same state by using our QAS method instead. 
As QAS does not adjust the parameters $\boldsymbol{\theta}$, and instead only varies the classical parameter $\alpha(t)$ of the linear combination of states ansatz, we can circumvent the barren plateau problem.
We now evolve this deep circuit as initial state with the Hamiltonian  $H_\text{bp}=\sigma_1^z\sigma_2^z$ as chosen by Ref.\cite{mcclean2018barren}. We measure the overlaps Eqs.\eqref{eq:Overlap_D},\eqref{eq:Overlap_E} and solve Eq.\eqref{eq:Evolution_Frenkel_Complex}. }
We find that we can already reproduce the time evolution of this state for the first moment expansion $K=1$.

\begin{figure}[htbp]
	\centering
	\subfigimg[width=0.24\textwidth]{a}{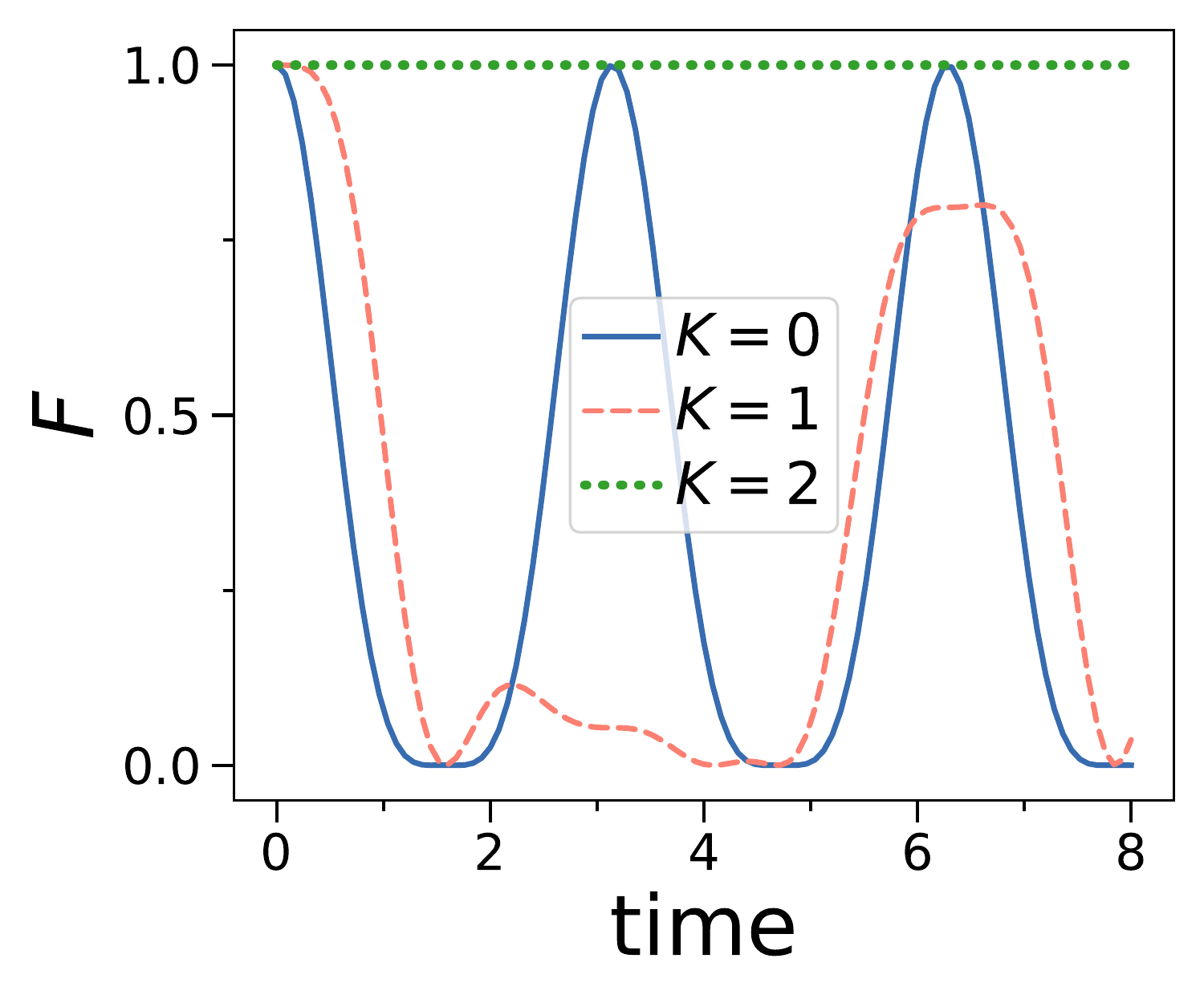}\hfill
	\subfigimg[width=0.24\textwidth]{b}{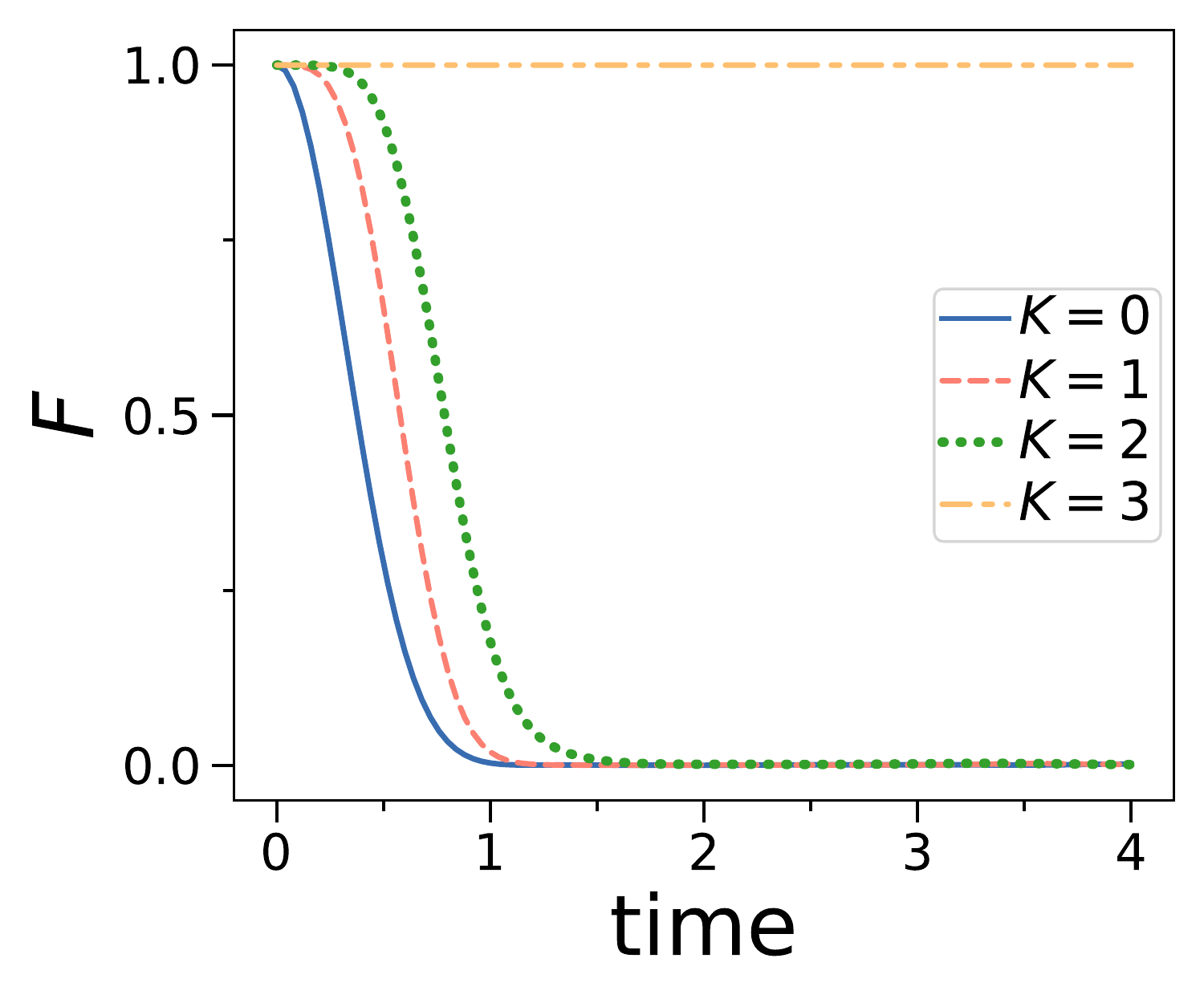}
	\caption{Fidelity $F=\abs{\braket{\Psi_\text{exact}(t)}{\phi(\alpha(t)}}^2$ against evolution time. Initial state is a deep random circuit of $N=10$ qubits and depth $d=200$ using same Ansatz as in Fig.\ref{Examples}b for varying moment expansion $K$. We evolve with two different Hamiltonians. \idg{a} Fermionic Hamiltonian $H=\frac{1}{2}(X[\otimes_{i=2}^{N-1}Z]X+Y[\otimes_{i=2}^{N-1}Z]Y)$ with Jordan-Wigner string $\otimes_{i=2}^{N-1}Z$, which represents tunneling of fermionic particles between first and last site of a chain. \idg{b} Ising Hamiltonian Eq.~\eqref{eq:Ising} with $J=1$ and $h=1$. }
	\label{ExamplesFid}
\end{figure}

Next, we discuss two further examples for the same variational Ansatz for different Hamiltonians in Fig.\ref{ExamplesFid}.
We evolve in Fig.\ref{ExamplesFid}a this Ansatz with the Hamiltonian $H=\frac{1}{2}(\sigma^x[\otimes_{i=2}^{N-1}\sigma^z]\sigma^x+\sigma^y[\otimes_{i=2}^{N-1}\sigma^z]\sigma^y)$. After applying the Jordan-Wigner transformation~\cite{jordan1993paulische}, this Hamiltonian corresponds to a system of fermions tunneling from the first to the last site in a lattice system of length $N$ with $H=c_1^\dagger c_N+c_N^\dagger c_1$, where $c_i^\dagger$ ($c_i$) is the fermionic creation (annihilation) operator acting on site $i$. Evolving this Hamiltonian with naive implementations of Trotter can be difficult as it requires implementing a $N$-qubit rotation. We show fidelity of QAS  with the exact evolution $F=\abs{\braket{\Psi_\text{exact}(t)}{\phi(\alpha(t)}}^2$. We find optimal fidelity for moment $K=2$.
In Fig.\ref{ExamplesFid}b we show the evolution of an exemplary many-body Hamiltonian, the Ising model. The Hamiltonian is given by 
\begin{equation}\label{eq:Ising}
H_\text{ising}=\frac{J}{2}\sum_{i=1}^N\sigma^x_i\sigma^x_{i+1}+\frac{h}{2}\sum_{i=1}^N\sigma^z_i\,,
\end{equation}
which describes a $N$ spin system with nearest-neighbor coupled spins with amplitude $J$ and an applied transverse magnetic field $h$. We find increasing fidelity with increasing $K$. \revA{For $K=3$, the $K$-moment states cover the full dynamics of the problem and thus we achieve unit fidelity.}

\begin{figure}[htbp]
	\centering
	\subfigimg[width=0.24\textwidth]{}{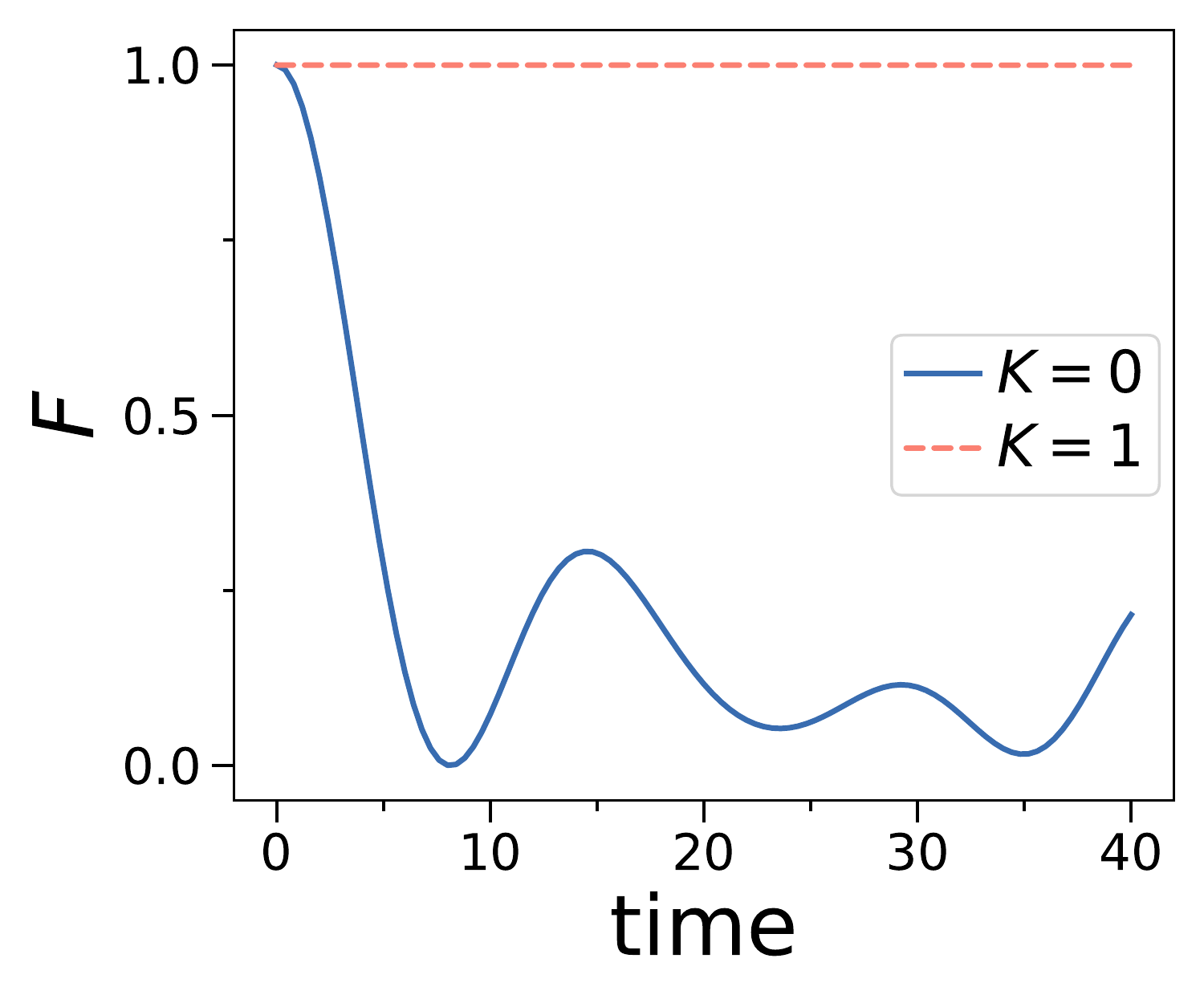}
	\caption{LiH molecule evolution for a perturbed excited state. The initial state is a superposition state consisting of the 10 eigenstates with lowest energy of the LiH molecule. We show fidelity $F=\abs{\braket{\Psi_\text{exact}(t)}{\phi(\alpha(t)}}^2$ against evolution time.  }
	\label{ExamplesLih}
\end{figure}

Now, we discuss QAS for dynamical simulation of a chemistry problem. In Fig.\ref{ExamplesLih}, we show the evolution of an excited state of a LiH molecule. The LiH molecule Hamiltonian is mapped to 6 qubits using STO-3G basis using methods of~\cite{mcardle2020quantum}, resulting in a Hamiltonian with 174 terms.  We find that for the $K=1$ moment, the dynamics can be fully reproduced with QAS. 

\begin{figure}
    \centering
    \includegraphics[width=0.35\textwidth]{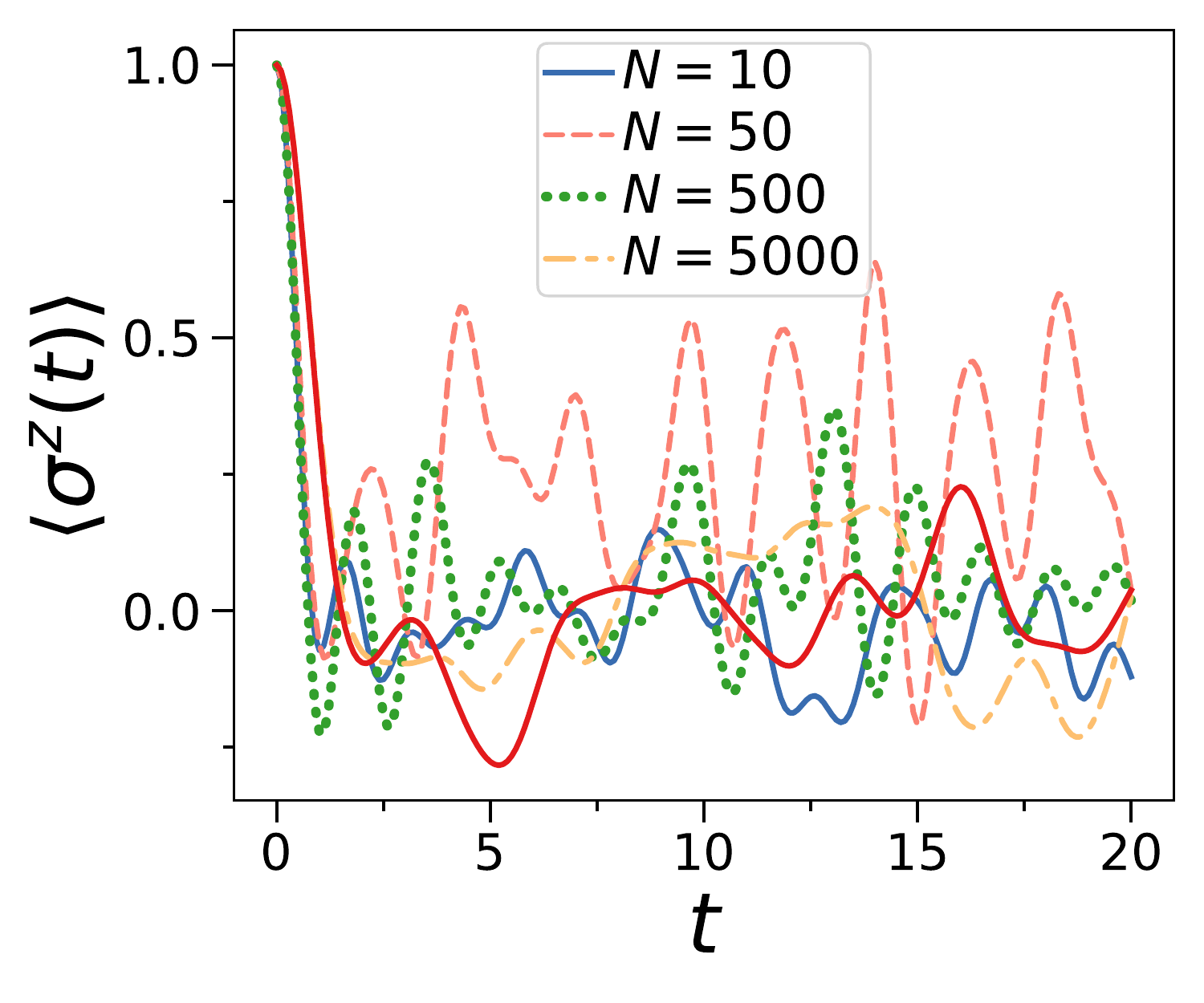}
    \caption{Simulation of a Hamiltonian of multi-body Pauli strings for a large number of qubits $N$. Our Hamiltonian consists of $r$ random Pauli strings $H=\sum_{i=1}^r P_i$, where we define the $N$-body Pauli strings $P_i=\otimes_{j=1}^N \boldsymbol{\sigma}_j$ which consist of $N$ tensored Pauli operators $\boldsymbol{\sigma}_j\in\{I,\sigma^x,\sigma^y,\sigma^z\}$.  
    Initially, we start with the $N$-qubit product state with all zeros. The ansatz states consists of $2^r=512$ states. }
    \label{fig:r_evolve}
\end{figure}
\revB{In Fig.\ref{fig:r_evolve}, we evolve a many-body Hamiltonian $H=\sum_{i=1}^r P_i$ that consists of $r$ randomly chosen $N$-body Pauli strings $P_i=\otimes_{j=1}^N \boldsymbol{\sigma}_j$, where $\boldsymbol{\sigma}_j\in\{I,\sigma^x,\sigma^y,\sigma^z\}$. 
The cumulative $K$-moment states capture at order $K=r$ the full dynamics. We can simulate the dynamics for thousands of qubits $N$.
We show the simulation of the dynamics of an initial product state using a classical computer. For a general entangled initial state, the overlaps required for QAS are intractable for classical computers and one  requires a quantum computer to simulate the dynamics.}

\begin{figure}[htbp]
	\centering
	\subfigimg[width=0.28\textwidth]{}{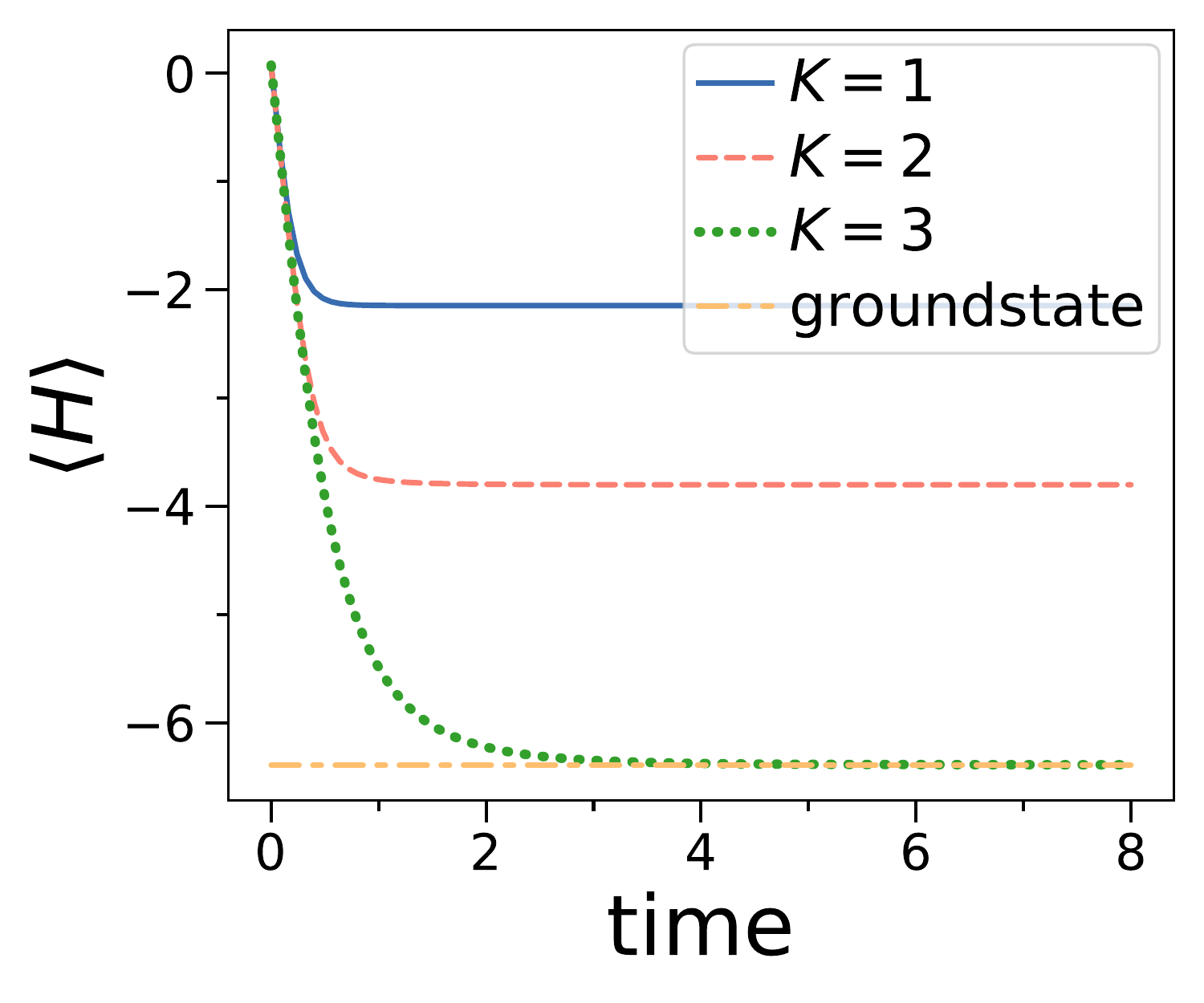}
	\caption{Imaginary time evolution of energy $\langle H(t)\rangle$ for QAS. Simulation with $N=10$ qubits ($J=1$, $h=1$) and randomized deep variational circuit as initial Ansatz state.  }
	\label{ExampleImag}
\end{figure}

Finally, we discuss imaginary time evolution with QAS. We defer the details to Appendix~\ref{sec: ITE}. In Fig.\ref{ExampleImag}, we show the energy $\langle H(t)\rangle$ of the imaginary time evolution of the Ising model for the deep variational Ansatz introduced above. With longer evolution time, the energy of the quantum state decreases, until it reaches a minimal value. The found minimal energy decreases with increasing $K$, reaching the exact ground state for $K=3$.

We show further examples for the simulated dynamics for a quenched many-body system in Appendix~\ref{sec:quench}.

%\medskip
%{\noindent {\em Discussion and Conclusion---}}
\section{Discussion and Conclusion}
In this work, we provided a hybrid quantum-classical algorithm for
simulating the dynamical evolution of a quantum system. Without loss
of generality, we assume the Hamiltonian to be a linear combination
of unitaries. Our algorithm proceeds in three steps and does
not mandate any classical-quantum feedback loop.  For algorithms with feedback loops, the quantum computer has to wait until the classical computer has executed its task. Most of the currently available quantum computers are accessed in a queue fashion and the feedback loop yields the whole process extremely slow, as the waiting time in the queue can be a few hours. The absence of the feedback loop renders our protocol exceptionally  faster than its variational alternatives.
The first step of
the algorithm involves the selection of the Ansatz. We consider our Ansatz
to be a linear combination of quantum states, where the quantum states
belong to the set of cumulative $K$-moment states, constructed using the unitaries defining the Hamiltonian. The combination
coefficients are complex numbers in general. For the choice of the Ansatz in the first step, the second step
involves the computation of two overlap matrices on the quantum computer.
This step can be performed efficiently on existing quantum computers
without the requirement of any complicated measurements by measuring Pauli strings.
Once the overlap matrices have been computed, the third (and final) step of the QAS algorithm involves solving the differential equation and updating the
parameters corresponding to Eq.~\eqref{eq:Evolution_Frenkel_Complex}
and \eqref{eq:Time_evolution_alpha_1-1} on a classical computer. If
the desired accuracy has not been attained, one can re-run the whole
algorithm for a different choice of $K$.

\revA{It is in general difficult for classical computers to calculate quantum states. Here, the advantage of our algorithm comes into play. 
With a quantum computer, we can prepare quantum states out of reach for classical computers. Then, we measure overlaps via sampling from the quantum computers, which is in general a difficult task for classical computers, including tensor network methods. 
Above statement is bolstered by the Quantum Threshold Assumption (QUATH)~\cite{aaronson2016complexity} by Aaronson and Chen. It states that there is no polynomial-time classical algorithm which takes as input a random circuit $C$ and can decide with success probability at least $\frac{1}{2}+\Omega\left(\frac{1}{2^n}\right)$ that whether $\vert\langle0^n \vert C \vert0^n \rangle\vert^2$ is greater than the median of $\vert\langle0^n \vert C \vert x^n \rangle\vert^2$ taken over all bitstrings $x^n$.
Finally, the task of integrating the time evolution using the overlaps is left to the classical computer, as this task can be efficiently performed there. 
Our algorithm solves the task of quantum simulation by leaving classically easy tasks to the classical computer, and distributing only the classically difficult tasks to the quantum computer. This way, we use the resources available in the NISQ era prudently. }

Apart from Dirac and Frenkel variational principle, there are two other variational principles, which can be used for variational simulation of dynamics: the McLachlan variational principle \cite{mclachlan1964variational} and the time-dependent variational method principle \cite{kramer1980geometry,broeckhove1988equivalence}. For an Ansatz with complex adjustable parameters, all three variational principles lead to same dynamical evolution \cite{yuan2019theory}. We also discussed the extension
of our algorithm to imaginary time evolution (see Appendix~\ref{sec: ITE} for details). It is noteworthy to stress that the quantum states defining the Ansatz are fixed, and only the variational parameters are (classically) updated. The algorithm does not mandate the computation of gradients using the quantum computer and thus circumvents the barren plateau problem by construction. 
Our approach of linear combination of quantum states could offer an advantage even at the level of initial state preparation.
There exists  cases where the initial state is difficult to prepare, but is respresentable as linear combination of $K$-moment states for some efficiently preparable zero-moment state. Such cases though intractable via variational algorithms such as VQS, are easily simulable with our method. Our algorithm can easily subsume VQS using the following Ansatz
\begin{equation}
\vert\phi\left(\boldsymbol{\alpha}(t),\boldsymbol{\theta}\right)\rangle=\sum_{i=0}^{m-1}\alpha_{i}(t)\vert\psi_{i}\left(\theta_{i}\right)\rangle,\label{eq:VQS_QAS_ANsatz}
\end{equation}
where $\alpha_{i}\in\mathbb{C}$ and $\theta_{i}\in\mathbb{R}.$ 

Since we unlock a fresh paradigm, there are many open questions for
further investigation. In future, one could study the extension of our
algorithm for simulating open system dynamics, in the presence of noise as well as for Gibbs state preparation. Extending our algorithm
to simulate generalized time evolution is another exciting direction~\cite{endo2020variational}.
It would be interesting to provide complexity-theoretic guarantees as well as analyze the error in-depth. A proper study of the
Ansatz construction is another exciting direction. 
We believe that an in-depth study of our algorithm will lead to the design of quantum-inspired classical algorithms for simulating the dynamics of quantum systems.

%\medskip
%{\noindent {\em Acknowledgements---}} 
\begin{acknowledgements}
We are grateful to the
National Research Foundation and the Ministry of Education, Singapore
for financial support. We thank Jonathan for interesting discussions.
\end{acknowledgements}

\bibliographystyle{apsrev4-1}
\bibliography{QAS}

\appendix 

\newpage

\section{Error Analysis} \label{sec: Error}
The error accumulated during the time of evolution can be calculated
by estimating the distance between the true evolution and the evolution
of the Ansatz \cite{yuan2019theory}. This is given by 
\begin{equation}
\epsilon_{t}=\left\Vert \left(\frac{d}{dt}+\iota H\right)\vert\phi\left(\boldsymbol{\alpha}(t)\right)\rangle\right\Vert ^{2},\label{eq:error_evolution}
\end{equation}
where $\left\Vert \vert\xi\rangle\right\Vert ^{2}=\langle\xi\vert\xi\rangle.$
Using Eq.\ref{eq:error_evolution}, the error term $\epsilon_{t}$ is given
by
\begin{multline}
\epsilon_{t}=\sum_{i,j}\frac{\partial\langle\phi\left(\boldsymbol{\alpha}(t)\right)\vert}{\partial\alpha_{i}^*}\frac{\partial\vert\phi\left(\boldsymbol{\alpha}(t)\right)\rangle}{\partial\alpha_{j}}\dot{\alpha_{i}}^{\star}\dot{\alpha_{j}}+ \\ \iota\sum_{i}\frac{\partial\langle\phi\left(\boldsymbol{\alpha}(t)\right)\vert}{\partial\alpha_{i}^*}H\vert\phi\left(\boldsymbol{\alpha}(t)\right)\rangle\dot{\alpha_{i}}^{\star} \\
- \iota\sum_{i}\langle\phi\left(\boldsymbol{\alpha}(t)\right)\vert H\frac{\partial\vert\phi\left(\boldsymbol{\alpha}(t)\right)\rangle}{\partial\alpha_{i}}\dot{\alpha_{i}}+ \\ \langle\phi\left(\boldsymbol{\alpha}(t)\right)\vert H^{2}\vert\phi\left(\boldsymbol{\alpha}(t)\right)\rangle.\label{eq:Error_evolution_1}
\end{multline}

\section{Realification} \label{sec: realify}
Recall that the Ansatz evolution equation is given by
\begin{equation}
\mathcal{E}\frac{\partial\boldsymbol{\alpha}(t)}{\partial t}=-\iota\mathcal{D}\boldsymbol{\alpha}(t),\label{eq:Evolution_Frenkel_Complex_Appendix}
\end{equation}
where $\alpha_{i}(t)\in\mathbb{C}$. Moreover, the overlap matrices
$\mathcal{E}$and $\mathcal{D}$ can be in general complex. We can
use realification to solve the above differential equation for $\boldsymbol{\alpha}(t)$. The idea of realification is to create a mapping from $\mathbb{C}^{n}$
to $\mathbb{R}^{2n}.$ The crux of the idea is the following real
matrix representation for $1$ and $i$,
\begin{equation}
1\leftrightarrow\left[\begin{array}{cc}
1 & 0\\
0 & 1
\end{array}\right]\equiv\mathbb{I}\label{eq:map_1}
\end{equation}
and
\begin{equation}
i\leftrightarrow\left[\begin{array}{cc}
0 & -1\\
1 & 0
\end{array}\right]\equiv I.\label{eq:map_iota}
\end{equation}
It can be seen that $I^{2}=-\mathbb{I},$which mimics $i^{2}=-1.$
Let $\mathcal{E_{R}}$, $\mathcal{D_{R}}$ and $\mathcal{E_{I}}$,$\mathcal{D_{I}}$
be the real and imaginary parts of the overlap matrices $\mathcal{E}$ and
$\mathcal{D}.$ Let $\boldsymbol{\alpha}_{R}(t)$ and $\boldsymbol{\alpha}_{I}(t)$
be the real and imaginary parts of $\boldsymbol{\alpha}(t).$
Using the mappings in \ref{eq:map_1} and \ref{eq:map_iota}, one can obtain the following representation for 
\begin{equation}
\mathcal{E}=\left[\begin{array}{cc}
\mathcal{E_{R}} & \mathcal{-E_{I}}\\
\mathcal{E_{I}} & \mathcal{E_{R}}
\end{array}\right],\label{eq:E_realify}
\end{equation}

\begin{equation}
\mathcal{D}=\left[\begin{array}{cc}
\mathcal{D_{R}} & \mathcal{-D_{I}}\\
\mathcal{D_{I}} & \mathcal{D_{R}}
\end{array}\right],\label{eq:D_realify}
\end{equation}
and 

\begin{equation}
\boldsymbol{\alpha}(t)=\left[\begin{array}{cc}
\dot{\boldsymbol{\alpha}}_{R}(t) & -\dot{\boldsymbol{\alpha}}_{I}(t)\\
\dot{\boldsymbol{\alpha}}_{I}(t) & \dot{\boldsymbol{\alpha}}_{R}(t)
\end{array}\right].\label{eq:Alpha_realify}
\end{equation}
Using the above representation, one trivially gets the following realified evolution
equation.
\begin{equation}
\left[\begin{array}{cc}
\mathcal{E_{R}} & \mathcal{-E_{I}}\\
\mathcal{E_{I}} & \mathcal{E_{R}}
\end{array}\right]\left[\begin{array}{c}
\dot{\boldsymbol{\alpha}}_{R}(t)\\
\dot{\boldsymbol{\alpha}}_{I}(t)
\end{array}\right]=\left[\begin{array}{cc}
\mathcal{D_{R}} & \mathcal{-D_{I}}\\
\mathcal{D_{I}} & \mathcal{D_{R}}
\end{array}\right]\left[\begin{array}{c}
\boldsymbol{\alpha}_{R}(t)\\
\boldsymbol{\alpha}_{I}(t)
\end{array}\right].\label{eq:realified_evolution_Frankel}
\end{equation}
\section{Imaginary Time Evolution} \label{sec: ITE}
Substituting $t$ with $-\iota\tau$ in the Schrödinger equation
we get
\begin{equation}
\frac{d\vert\psi(\tau)\rangle}{dt}=-\left(H-\left\langle H\right\rangle \right)\vert\psi(\tau)\rangle,\label{eq:ITE}
\end{equation}
where $\left\langle H\right\rangle =\langle\psi(\tau)\vert H\vert\psi(\tau)\rangle.$
The presence of $\left\langle H\right\rangle $ preserves the norm
of $\vert\psi(\tau)\rangle.$ Let us consider the Ansatz state as linear combination of $m$ quantum states $\left\{ \vert\psi_{i}\rangle\right\}_{i}$
\begin{equation}
\vert\phi\left(\boldsymbol{\alpha}(\tau)\right)\rangle=\sum_{i=0}^{m-1}\alpha_{i}(\tau)\vert\psi_{i}\rangle,\label{eq:Ansatz-ITE}
\end{equation}
for $\alpha_{i}(\tau)\in\mathbb{C}.$ Using Dirac and Frenkel variational
principle, we get

\begin{equation}
\langle\delta\phi\left(\boldsymbol{\alpha}(\tau)\right)\vert\left(\frac{d}{d\tau}+H-\left\langle H\right\rangle \right)\vert\phi\left(\boldsymbol{\alpha}(\tau)\right)\rangle=0,\label{eq:Frenkel-ITE}
\end{equation}
where 
\begin{equation}
\langle\delta\phi\left(\boldsymbol{\alpha}(\tau)\right)\vert=\sum_{i}\frac{\partial\langle\phi\left(\boldsymbol{\alpha}(\tau)\right)\vert}{\partial\alpha_{i}^*}\delta\alpha_{i}^*.\label{eq:Frenkel_variation-ITE}
\end{equation}

The evolution of parameters is given by

\begin{equation}
\sum_{j}\mathcal{E}_{i,j}\dot{\boldsymbol{\alpha}_{j}}=-\mathcal{G}_{i},\label{eq:Evolution_Frenkel_ITE}
\end{equation}
where $\mathcal{E}$ and $\mathcal{G}$ are 

\begin{equation}
\mathcal{E}_{i,j}=\frac{\partial\langle\phi\left(\boldsymbol{\alpha}(\tau)\right)\vert}{\partial\alpha_{i}^*}\frac{\partial\vert\phi\left(\boldsymbol{\alpha}(\tau)\right)\rangle}{\partial\alpha_{j}},\label{eq:Overlap_E-1}
\end{equation}

\begin{equation}
\mathcal{G}_{i}=\sum_{j}\alpha_{j}(\tau)\frac{\partial\langle\phi\left(\boldsymbol{\alpha}(\tau)\right)\vert}{\partial\alpha_{i}^*}\left(H-\left\langle H\right\rangle \right)\vert\psi_{j}\rangle.\label{eq:Overlap_C-1}
\end{equation}
Further simplification leads to 
\[
\mathcal{G}_{i}=\left(\sum_{j}\alpha_{j}(\tau)\langle\psi_{i}\vert H\vert\psi_{j}\rangle\right)-\left\langle H\right\rangle \left(\sum_{j}\alpha_{j}(\tau)\langle\psi_{i}\vert\psi_{j}\rangle\right)
\]

\begin{equation}
\implies \mathcal{G}_{i}=\sum_{j}\mathcal{D}_{i,j}\alpha_{j}(\tau)-\left\langle H\right\rangle \sum_{j}\mathcal{E}_{i,j}\alpha_{j}(\tau).\label{eq:Overlap_G_ITE_2}
\end{equation}

\section{Quantum Quench} \label{sec:quench}
\begin{figure}[htbp]
	\centering
	\subfigimg[width=0.24\textwidth]{a}{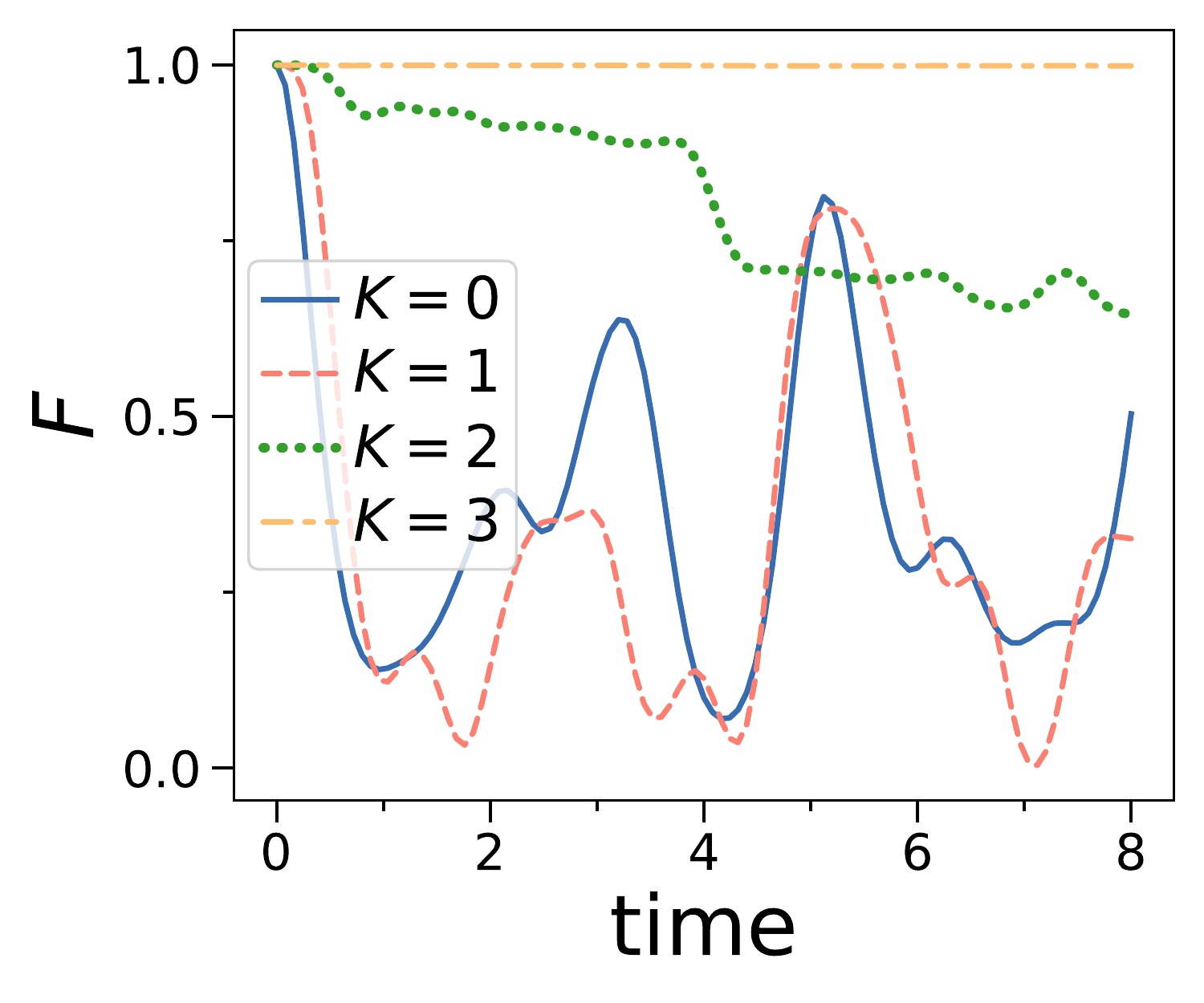}\hfill
	\subfigimg[width=0.24\textwidth]{b}{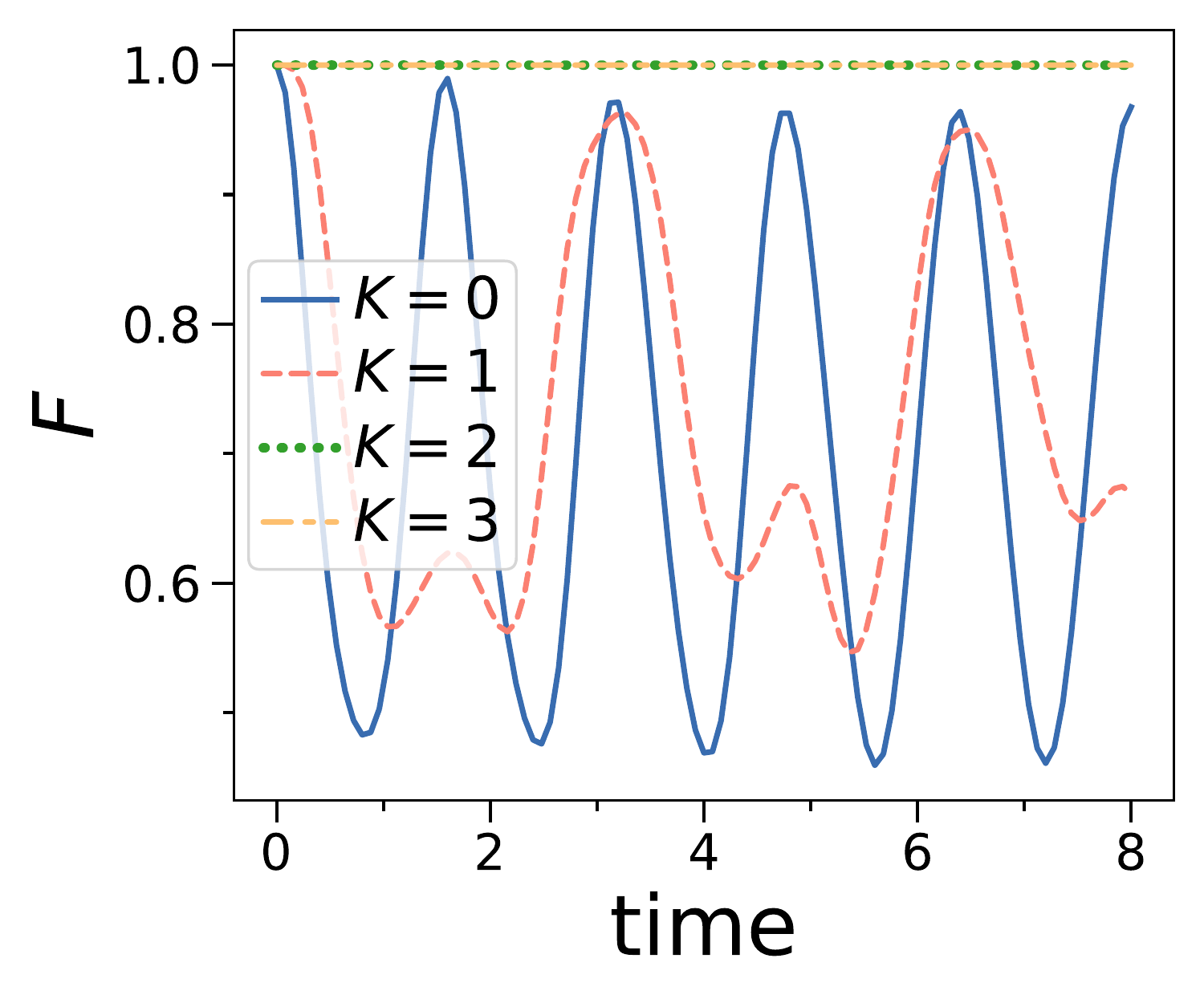}
	\caption{Many-body quench against time with fidelity $F=\abs{\braket{\Psi_\text{exact}(t)}{\phi(\alpha(t)}}^2$ in respect to exact dynamics for varying moment expansion $K$. \idg{a} Ising Hamiltonian. \idg{b} Heisenberg model. In both cases, we prepare the initial ground state of Hamiltonian with $J=1$, $h=0.5$, and then quench the Hamiltonian parameter instantaneously to $h=2$. Number of qubits is $N=8$.  }
	\label{ExamplesQuench}
\end{figure}

We now give further numerical examples for QAS.
We demonstrate quench dynamics for many-body systems. Here, the system is prepared in the ground state of a Hamiltonian. Then,  to generate the dynamics,  the parameters of the Hamiltonian are instantaneously changed to another value. We show two examples for such dynamics in Fig.\ref{ExamplesQuench}. 
We study the Ising Hamiltonian given by 
\begin{equation}
H_\text{ising}=\frac{J}{2}\sum_{i=1}^N\sigma^x_i\sigma^x_{i+1}-\frac{h}{2}\sum_{i=1}^N\sigma^z_i\,,
\end{equation}
and the Heisenberg model
\begin{equation}
H_\text{XXZ}=\frac{1}{2}\sum_{i=1}^N(\sigma^x_i\sigma^x_{i+1}+\sigma^y_i\sigma^y_{i+1} + \Delta\sigma^z_i\sigma^z_{i+1})\,\,
\end{equation}
which describe $N$ spin systems with nearest-neighbor coupled spins with parameters $J$ and $h$. 
We find that for the Ising Hamiltonian (Fig.\ref{ExamplesQuench}a), the evolution becomes more accurate with increasing $K$, reaching optimal fidelity for $K=4$.
In Fig.\ref{ExamplesQuench}b, we study the Heisenberg model. We find that we converge to optimal fidelity for $K\ge3$.

\section{Measurements} \label{sec: Measurements}
We reiterate here the content from reference \cite{bharti2020iterative}. For the purposes of this section, a unitary $U$ will be referred $k$-local if it acts non-trivially on at most $k$ qubits. We assume $k=\mathcal{O}(poly(log(N))$. The moment-$K$ Ansatz, for some positive integer $K$, in QAS algorithm requires computation of matrix elements
of the form $\langle\psi^\vert U\vert\psi\rangle,$
where $U$ is a k-local unitary matrix (since $U$ is product of at most $2K+1$
k-local unitary matrices). By invoking the following result from Mitarai \textit{et.
al.} \cite{mitarai2019methodology}, we guarantee an efficient
computation of the overlap matrices on a quantum computer without the use of the
Hadamard test. 
\begin{fact} \cite{mitarai2019methodology} \label{fact: hadamard}
Let k be an integer such that $k=\mathcal{O}(poly(log(N))$, where
$N$ is the number of qubits and $\ket{\psi}$ be an $N$-qubit quantum state. For any k-local quantum gate $U,$ it
is possible to estimate $\langle\psi \vert U\vert \psi \rangle$
up to precision $\epsilon$ in time $\mathcal{O}\left(\nicefrac{k^{2}2^{k}}{\epsilon^{2}}\right)$
without the use of the Hadamard test, with classical preprocessing
of time $\mathcal{O}(poly(logN)).$
\end{fact}
We proceed to discuss the methodology suggested in \citep{mitarai2019methodology}
to calculate the required matrix elements. For detailed analysis,
refer to \citep{mitarai2019methodology}. Since $U$ is a $k$-local
unitary, it can be decomposed as $U=\otimes_{q=1}^{Q}U_{q}$ such
that $U_{q}$ acts on $k_{q}$ qubits. Clearly, $U_{q}$ is a $2^{k_{q}}\times2^{k_{q}}$
matrix. Suppose the eigenvalues of $U_{q}$ are $\left\{ \exp\left(i\phi_{q,m}\right)\right\} _{m=0}^{2^{k_{q}}-1}.$
Using the integers $m_{q}=0,\cdots,2^{k_{q}}-1,$ let us denote the
computational basis of each subsystem by $\vert m_{q}\rangle.$ We
diagonalize each $U_{q}$ and obtain some unitary matrix $V_{q}$
such that $U_{q}=V_{q}^{\dagger}T_{q}V_{q},$ where $T_{q}=\sum_{m=0}^{2^{k_{q}}-1}\exp\left(i\phi_{q,m}\right).$
Since $k=\mathcal{O}(poly(log(N))$, the aforementioned diagonalized
can be performed in polynomial time. A simple calculation gives,
\begin{multline}
 \langle\psi\vert U\vert\psi\rangle=\sum_{m_{1}=0}^{2^{k_{1}}-1}\cdots\sum_{m_{Q}=0}^{2^{k_{Q}}-1}\left(\prod_{q=1}^{Q}\exp\left(i\phi_{q,m_{q}}\right)\right)\\
\times\left|\left(\otimes_{q=1}^{Q}\langle m_{q}\vert\right)\left(\otimes_{q=1}^{Q}V_{q}\right)\vert \psi \rangle\right|^{2}.\label{eq:matrix_elements_evaluation}
\end{multline}
Thus, one can evaluate  $\langle\psi^\vert U\vert\psi\rangle$
by calculating the probability of getting $\otimes_{q=1}^{Q}\vert m_{q}\rangle$
from the measurement of $\left(\otimes_{q=1}^{Q}V_{q}\right)\vert \psi \rangle$
in the computational basis.
\begin{figure}[htbp]
\includegraphics[scale=0.2]{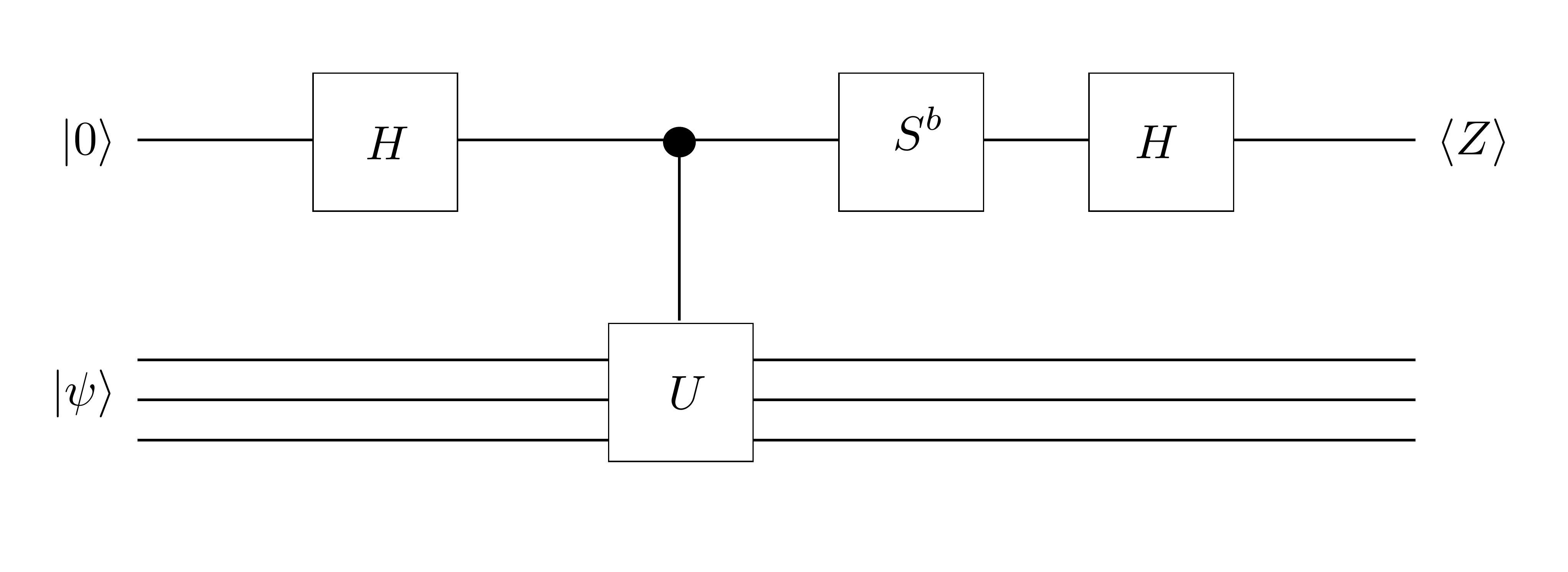}
\caption{The above figure shows a simple Hadamard test circuit for measuring
the real and imaginary part of $\langle \psi \vert U\vert \psi \rangle$
for any arbitrary $n$ qubit unitary $U$. The Hadamard gate has been
represented by $H$ and $S$ represents $e^{-\nicefrac{\iota\pi Z}{4}}.$
When $b=0,$ we get $\left\langle Z\right\rangle =\text{Re}\left\langle \psi \vert U\vert \psi \right\rangle .$
For $b=1,$ we get $\left\langle Z\right\rangle =\text{Im}\left\langle \psi\vert U\vert \psi \right\rangle .$
Since implementing controlled-unitaries is challenging in the NISQ
era, the use of Hadamard test as a subroutine is highly discouraged
while designing NISQ-friendly quantum algorithms. Using results from
\cite{mitarai2019methodology}, we guarantee an efficient computation of the overlap matrices required in the QAS algorithm on a quantum computer without the use
of the Hadamard test.}
\label{fig: Hadamard}
\end{figure}

If the unitaries defining the Hamiltonian are Pauli strings over $N$ qubits, one can use the following Lemma \cite{huang2019near} to provide an estimate of the number of measurements needed to achieve a given desired accuracy. 
\begin{lem} \cite{huang2019near}
Let $\epsilon>0$ and $P_{q}$ be a  Pauli string over $N$
qubits. Let multiple copies of an arbitrary $N$-qubit quantum state
$\vert\psi\rangle$ be given. The expectation value $\langle\psi\vert P_{q}\vert\psi\rangle$
can be determined to additive accuracy $\epsilon$ with failure probability
at most $\delta$ using $\mathcal{O}\left(\frac{1}{\epsilon^{2}}\log\left(\frac{1}{\delta}\right)\right)$
copies of $\vert\psi\rangle.$
\end{lem}
We now discuss how to measure the matrix elements of $\mathcal{D}$ and $\mathcal{E}$ for the special case where the components $U_i$ of the Hamiltonian $H=\sum_i\beta_i U_i$ are Pauli strings $P_q=\bigotimes_{j=1}^N\gamma_j$, with $\gamma\in \{\mathbb{1},\sigma^x,\sigma^y,\sigma^z\}$. 
For a $K$-moment state expansion of Ansatz $\ket{\Psi}$, the elements to calculate are
\begin{align*}
\mathcal{D}_{n,m}&=\sum_i\beta_i\bra{\Psi}U_{n_1}^\dagger\dots U_{n_K}^\dagger U_i U_{m_K}\dots U_{m_1}\ket{\Psi}\\ \mathcal{E}_{n,m}&=\bra{\Psi}U_{n_1}^\dagger\dots U_{n_K}^\dagger U_{m_K}\dots U_{m_1}\ket{\Psi}. \numberthis\label{Eq:Pauli_matrix}
\end{align*}
Now, each overlap element is a product of a set of some Pauli strings $P_{q}$ to be measured on state $\ket{\Psi}$, with $\bra{\Psi}\prod_{q} P_{q}\ket{\Psi}$. 
The product rule of Pauli operators states that $\sigma^i\sigma^j=\delta_{ij}\mathbb{1}+i\epsilon_{ijk}\sigma^k$, where $\sigma^1=\sigma^x,\, \sigma^2=\sigma^y,\, \sigma^3=\sigma^z$, $\delta_{ij}$ is the Kronecker delta and $\epsilon_{ijk}$ the Levi-Civita symbol. Thus, a product of two Pauli strings $P_q P_p=a P_s$ is again a Pauli string $P_s$, with a prefactor $a \in \{+1,-1,+i,-i\}$. 
To calculate the matrix elements on the quantum computer, first one has to evaluate which Pauli string corresponds to the product of unitaries in Eq.\ref{Eq:Pauli_matrix} and the corresponding prefactor. Then, the expectation value of the resulting Pauli string is measured for the Ansatz state $\bra{\Psi}P_q\ket{\Psi}$. This observable is a hermitian operator, and can be easily measured by rotating each qubit into the computational basis corresponding to the Pauli operator. Finally, the expectation value of the measurement is multiplied with the prefactor $a$.
\newpage
\end{document}